# Full-separability and bi-separability of qubits using Bell operators, partial transpose, witnesses and explicit (full/bi) separability

Y. Ben-Aryeh and A. Mann

Physics Department, Technion-Israel Institute of Technology, Haifa 32000

E-mails: phr65yb@physics.technion.ac.il ; ady@physics.technion.ac.il

**Abstract:** Using the Hilbert-Schmidt ($HS$) decomposition we suggest new possible choices of Bell operators and entanglement witnesses ($EW$) for n (>2) qubits systems for (full/bi) separability. The latter give upper bounds for (full/bi) separability. Also using the $HS$ decomposition, we find explicitly (full/bi) separable forms for some qubits states which give lower bounds for (full/bi) separability. When the lower bounds and upper bounds coincide it means that the $EW$ is optimal. In the case of full separability, the positive transpose method can sometimes give optimal results. As concrete examples, we give results for the $GHZ(3)$, $W(3)$ and cluster $Cl(4)$ states.

**Condensed paper title:** Full-separability and bi-separability of qubits

## 1. Introduction

The importance of entanglement in quantum mechanics is enormous. Entanglement of qubits systems is at the core of the quantum computation field. There is much interest in quantum entangled states due to various potential applications that use the quantum properties of such states. The most famous application is the use of quantum systems for a new generation of computers that will be based on principles of quantum computation (QC). Therefore it is of utmost importance to quantify entanglement in such systems and to have a definite criterion when such systems are separable.

There are various methods to determine entanglement of quantum states. For 2 qubits the Peres-Horodecki (PH) [1, 2] partial transpose ($PT$) criterion is necessary and sufficient for entanglement. For more than 2 qubits there is no such criterion. Also for more than 2 qubits there are various possibilities for entanglement, e.g., for 3 qubits there is full separability, bi-separability and genuine entanglement [3, 4].

A pure state is fully separable ($fs$) if it is of the form $\left|\psi^{fs}\right\rangle = |a\rangle \otimes |b\rangle \otimes |c\rangle \otimes \cdots$ . A density matrix is $fs$ if it can be written as [1]:

$$\rho^{fs} = \sum_k p_k \left|\psi_k{}^{fs}\right\rangle\left\langle\psi_k{}^{fs}\right| \quad ; \quad p_k \geq 0 \quad , \sum_k p_k = 1 \quad . \tag{1}$$

For n-qubits state explicit expression for full-separabiity can be given as

$$\rho^{fs}{}_{A,B,C\cdots} = \sum_j p_j \rho_A{}^{(j)} \otimes \rho_B{}^{(j)} \otimes \rho_C{}^{(j)} \cdots \; ; \quad p_j \geq 0 \quad , \; \sum_j p_j = 1 \qquad (2)$$

where $\rho_A{}^{(j)}, \rho_B{}^{(j)}, \rho_C{}^{(j)} \cdots$ are the density matrices of qubits $A, B, C, \cdots$ respectively.

A pure state $|\phi^{bs}\rangle$ is called bi-separable (bs) if it is separable under some partition, e. g., one partition might be given by

$$\phi^{bs} = \phi_{AB} \otimes \phi_C \qquad , \qquad (3)$$

where $\phi_{AB}$ is a pure state of qubits $A$ and B and $\phi_C$ is a pure state of qubit $C$

$$\phi_C = \begin{pmatrix} \cos\frac{\theta_C}{2} \\ e^{i\varphi_C} \sin\frac{\theta_C}{2} \end{pmatrix} \qquad . \qquad (4)$$

A density matrix is $bs$ [1, 2] if it can be written as

$$\rho^{bs} = \sum_k p_k \left|\phi_k{}^{bs}\right\rangle\left\langle\phi_k{}^{bs}\right| \quad ; \quad p_k \geq 0 \quad , \; \sum_k p_k = 1 \qquad . \qquad (5)$$

Explicit expressions for bi-separabity of 3-qubits (and more) have been analyzed in the literature [3-5]. A state is genuinely entangled if it is not $bs$.

A main obstacle for realizing quantum computational processes is the noise entering the quantum state and white noise is the simplest kind of such noise. In the present work we are interested in finding the condition for full separability and genuine entanglement of 3 and 4 qubits entangled states mixed with white noise. This problem is very important since if the state is fully separable no measurement on one qubit can affect the measurements of the other qubits (i. e. there is not any EPR effect). If the state is genuinely entangled i.e. not bi-separable under any partition, the measurement of any qubit can have a quantum mechanical effect on the others. We are interested in the following problem: Assuming entangled density matrix $\rho_{Ent}$ of n-qubits which is mixed with white noise, then $\rho_{Ent}$ becomes

$$\rho_{WN} = \left(\frac{1-p}{2^n}\right)(I)_A \otimes (I)_B \otimes (I)_C \cdots + p\rho_{Ent} \qquad . \qquad (6)$$

The subscript $WN$ denotes white noise with probability $1-p$ mixed with entangled state with density matrix $\rho_{Ent}$ with probability $p$.

One should take into account that the full Hilbert-Schmidt (HS) decomposition of n-qubits density matrix is given by

$$2^n \rho = (I)_A \otimes (I)_B \otimes (I)_C \cdots + \sum_{\substack{a,b,c\cdots=0 \\ (a,b,c\cdots)\neq(0,0,0\cdots)}}^{n} R_{a,b,c\cdots} (\sigma_a)_A \otimes (\sigma_b)_B \otimes (\sigma_c)_C \cdots \ ; \ \sigma_0 = I \quad , (7)$$

where the subscripts $A, B, C\cdots$ refer to the different qubits, the $\sigma's$ for $a,b,c \neq 0$ are the Pauli matrices and $\sigma_0$ is the unit $2\times2$ matrix. The HS parameter $R_{a,b,c\cdots}$ is given by $R_{a,b,c\cdots} = Tr(\rho\sigma_a\sigma_b\sigma_c\cdots)$ but in actual cases many of these parameters vanish. Partial-transform ($PT$) [6, 7] relative to qubit A ($PTA$) is obtained by inverting the sign of each term $R_{a,b,c\cdots}$ of Eq. (7) if and only if: $(\sigma_a)_A \equiv (\sigma_y)_A$ (i.e. , $PTA(\sigma_y)_A = -(\sigma_y)_A$ ). In a similar way we get the PT relative to qubit B ($PTB$ ) (i.e. , $PTB(\sigma_y)_B = -(\sigma_y)_B$ ) .

Using partial transpose ($PT$) for n-qubits states [1, 2] if we get a negative eigenvalue we conclude that this state is not fully separable (it may yet be bi-separable). But if we do not get a negative eigenvalue no conclusion is obtained on the non-separability of the state. Such situation occurs for example for a density matrix with maximally disordered subsystems (MDS) (i.e., a density matrix for which tracing over any subsystem gives the unit matrix of the remainder) [6]. Sufficient conditions and explicit expressions for separabilty of such states were analyzed in our previous work [5, 7, 8]. For a MDS density matrix with an odd number of qubits the eigenvalues of the PT matrix are the same as the original density matrix [7, 8] so it does not give information on separability. If PT does not give any information, common means to detect entanglement are entanglement witnesses. For n-qubits density matrix we can try to transform it into the form of Eq. (2) and if it works we conclude that the density matrix is fully separable. But if such attempt does not succeed no information is obtained, as there might be a different, better attempt. In our previous work we found explicit expressions for full separability of n-qubits systems [7, 8].

Bell operators and entanglement witnesses ($EW$) are very interesting since: a) Bell operators enable us to negate the possibility of a local–hidden-variables ($LHV$) model for a quantum state (e. g. [9-16]). b) $EW$ enable us to negate separability of a given quantum state (e.g. [17-29]). One should note that the use of $EW$ for non-full separability is different from the $EW$ for genuine entanglement [20]. The $EW$ which negates full separability has the same purpose as the $PT$. We use combinations of the above methods for getting optimal results. The use of $EW$ for non-full separability and the use of Bell operator to negate a $LHV$ model is explained in the next section.

## 2. Bell operators for n-qubits and $EW$ for non-full separability

$EW^{fs}$ is an entanglement witness for non-full separability of a density matrix $\rho$ if it satisfies two criterions given by

$$Tr\left(EW^{fs}\left|\psi^{fs}\right\rangle\left\langle\psi^{fs}\right|\right)\geq 0 \qquad (8)$$

where $\left|\psi^{fs}\right\rangle$ is any fully separable state and

$$Tr\left(EW^{fs}\rho\right)<0 \quad . \qquad (9)$$

We are interested in using the criterions of Eqs. (8, 9) for checking conditions for full separability and compare these results with those obtained by the $PT$ and/or with explicit full separability construction for the density matrix.

Let $\hat{O}$ represent either a Bell operator or $EW$. Given a density matrix $\rho$ the interesting quantity in both cases is given by $Tr(\hat{O}\rho)$. For the Bell case i. e. , for $\hat{O}=\hat{B}$, if this trace is larger than the classical bound of $\hat{B}$ there is no $LHV$ model for the state. Our method for choosing the Bell operator or $EW$ is based on the $HS$ decomposition for the given state. The $HS$ decomposition can be used for any operator $\hat{O}$. Note that for any operator $\hat{O}=\sum_{a,b,c,=0}^{n} O_{a,b,c,\cdots}(\sigma_a)_A\otimes(\sigma_b)_B\otimes(\sigma_c)_C\cdots$ we have

$$Tr(\hat{O}\rho)=\sum_{a,b,c} O_{a,b,c,\cdots}R_{a,b,c\cdots}\cdots \quad , \qquad (10)$$

where the parameters $R_{a,b,c,\cdots}$ were defined in Eq. (7).

Obviously the only terms that contribute to the expectation value $Tr(\hat{O}\rho)$ are those $HS$ terms that are common to $\hat{O}$ and $\rho$. Therefore $\hat{O}$ should be chosen to include in the $HS$ decomposition some (at least) of the $HS$ terms of $\rho$, not necessarily with the same coefficients. The appropriate Bell operator $\hat{B}$ consists of sums of products: $B_{a,b,c\cdots}(\sigma_a)_A\otimes(\sigma_b)_B\otimes(\sigma_c)_C\cdots$. We need to find the classical bound $\beta_{cl}$ for this expression (where for each $\sigma$ the value $+1$ or

-1 is assumed and the classical $\beta_{cl}$ is given by the maximal value for any possible summation of such values). For a given quantum state $\rho$ we calculate the quantum expectation value: $Tr(\hat{B}\rho) = \beta_{qu}$; if it breaks the classical bound i. e. $\beta_{qu} > \beta_{cl}$ then $\rho$ cannot be described by $LHV$ model.

An entanglement witness for full separability $EW^{fs}$ is

$$EW^{fs} = \alpha^{fs}(I)^n - G_{fs} \tag{11}$$

where for n-qubits state $I$ and $G_{fs}$ are the unit and a certain Hermitian operator, respectively, and $\alpha^{fs}$ is determined by Eq. (8). From Eqs. (8, 11) we get

$$\alpha^{fs} = \max\left\{Tr\left(\left|\psi^{fs}\right\rangle\left\langle\psi^{fs}\right|G_{fs}\right)\right\} = \max\left\langle\psi^{fs}\right|G_{fs}\left|\psi^{fs}\right\rangle \quad . \tag{12}$$

Mixing an entangled density matrix $\rho_{Ent}$ with white noise we obtain $\rho_{WN}$ according to Eq. (6). By increasing the parameter: $1-p$ in $\rho_{WN}$ we increase the component of white noise so that below a certain critical value $p_{crit}$, the mixed state $\rho_{WN}$ becomes fully separable. In order to find this critical value $p_{crit}$ we can use any witness $EW^{fs}$ which is defined by Eq. (11) and which satisfies the fully separability criterion of Eq. (8). Then by substituting $\rho_{WN}$ and $EW^{fs}$ in the criterion of Eq. (9) we get the condition for this $p_{crit}$

$$\begin{aligned}&\max Tr\left[EW^{fs}\rho_{WN}\right] = \\ &\max Tr\left[\left\{\alpha^{fs}I^{(n)} - \hat{G}_{fs}\right\}\cdot\left\{\left(\frac{1-p}{2^n}\right)(I)^{(n)} + p\rho_{Ent}\right\}\right] < 0\end{aligned} \quad . \tag{13}$$

The terms in the squared brackets of Eq. (13) can be rearranged as

$$\left[\alpha_{fs}(1-p) + \alpha_{fs}p - pTr\left(\hat{G}_{fs}\rho_{Ent}\right) - Tr\left[\frac{\hat{G}_{fs}(1-p)}{2^n}\right]\right] \quad . \tag{14}$$

Then we get:

$$\max Tr\left[EW^{fs}\rho_{WN}\right] = max\left[\alpha_{fs}(1-p) + \alpha_{fs}p - pTr\left(\hat{G}_{fs}\rho_{Ent}\right) - Tr\left[\frac{\hat{G}_{fs}(1-p)}{2^n}\right]\right] < 0 \tag{15}$$

The limiting critical value: ${p_{crit}}^{fs}$, above which the system is not fully separable, is given by

$$p_{crit}{}^{fs} = \frac{\alpha^{fs} - \frac{TrG_{fs}}{2^n}}{Tr\left(\rho_{Ent} G_{fs}\right) - \frac{TrG_{fs}}{2^n}} \quad . \tag{16}$$

It is a difficult task to find $G_{fs}$ which by calculating $\alpha_{fs}$ according to Eq. (12) and using Eq. (16) will give the optimal value $p_{wit}{}^{fs}$ which satisfies both conditions: above it is not fully separable according to Eq. (16) and below it is fully separable (according to the explicit calculations).

Any 3-qubits fully separable pure state density matrix $\rho_{fs}$ can be written as

$$\rho_{fs} = \frac{\left(I + \vec{l} \cdot \vec{\sigma}\right)_A}{2} \otimes \frac{\left(I + \vec{m} \cdot \vec{\sigma}\right)_B}{2} \otimes \frac{\left(I + \vec{n} \cdot \vec{\sigma}\right)_C}{2} \quad . \tag{17}$$

Here $I$ is the $2 \times 2$ unit matrix, the subscripts $A, B, C$ refer to the three qubits and $\vec{\sigma} = \sigma_x \hat{x} + \sigma_y \hat{y} + \sigma_z \hat{z}$ where $\sigma_x, \sigma_y, \sigma_z$ are the Pauli matrices. $\vec{l}, \vec{m}, \vec{n}$ are 3-dimensional vectors with a unit norm and with the components:

$$\begin{aligned}
&l_x = \sin(\theta)_A \cos(\varphi)_A \,;\, l_y = \sin(\theta)_A \sin(\varphi)_A \,;\; l_z = \cos(\theta)_A \,, \\
&m_x = \sin(\theta)_B \cos(\varphi)_B \,; m_y = \sin(\theta)_B \sin(\varphi)_B \,; m_z = \cos(\theta)_B \,, \\
&n_x = \sin(\theta)_C \cos(\varphi)_C \,; n_y = \sin(\theta)_C \sin(\varphi)_C \,; n_z = \cos(\theta)_C
\end{aligned} \quad . \tag{18}$$

Eq. (18) can be generalized for any n-qubits system where for 4-qubits system $\vec{l}, \vec{m}, \vec{n}$ will be changed to $\vec{l}, \vec{m}, \vec{n}, \vec{o}$ , etc. For 3-qubits the HS parameters of $\left|\psi^{fs}\right\rangle\left\langle\psi^{fs}\right|$ are given by

$$l_a m_b n_c = Tr\left[\left|\psi^{fs}\right\rangle\left\langle\psi^{fs}\right| (\sigma_a)_A \otimes (\sigma_b)_B \otimes (\sigma_c)_C\right] \quad ; \quad a, b, c = 0,1,2,3 \quad . \tag{19}$$

So, the HS decomposition of $\left|\psi^{fs}\right\rangle\left\langle\psi^{fs}\right|$ for 3-qubits is given by

$$2^3 \left|\psi^{fs}\right\rangle\left\langle\psi^{fs}\right| = \sum_{a,b,c=0}^{3} l_a m_b n_c (\sigma_a)_A \otimes (\sigma_b)_B \otimes (\sigma_c)_C \quad . \tag{20}$$

Without loss of generality we write

$$EW^{fs} = \alpha I^{(n)} - \hat{G}_{fs} \quad . \tag{21}$$

where $\hat{G}_{fs}$ is a certain operator (with $Tr(\hat{G}_{fs}) = 0$)) chosen as

$$\hat{G}_{fs} = \sum_{a',b',c'\cdots=0}^{3} G_{fs,a'.b',,c'\cdots} (\sigma_{a'})_A \otimes (\sigma_{b'})_B \otimes (\sigma_{c'})_C \cdots \quad ; \quad (a', b', c'\cdots) \neq (0,0,0\cdots) \quad . \tag{22}$$

This $\hat{G}_{fs}$ with the choice of coefficients: $G_{fs,a',b',c'...}$ leads to the value of $\alpha_{fs}$ and to the condition for non-separability by the criterions of Eqs, (8) and (9). In Eq. (8) we substitute $\hat{G}_{fs}$ by Eq. (22) and for $\left|\psi^{fs}\right\rangle\left\langle\psi^{fs}\right|$ by Eq. (20). Then we get:

$$\alpha^{fs} = \max Tr\left(\hat{G}_{fs}\left|\psi^{fs}\right\rangle\left\langle\psi^{fs}\right|\right) =$$
$$\max Tr\left(\sum_{a',b',c'...=0}^{3}\left(G_{fs,a',b',c'}\left(\sigma_{a'}\right)_A\otimes\left(\sigma_{b'}\right)_B\otimes\left(\sigma_{c'}\right)_C\right)\sum_{a,b,c=0}^{3}\left(l_a m_b n_c\left(\sigma_a\right)_A\otimes\left(\sigma_b\right)_B\otimes\left(\sigma_c\right)_C\right)/2^3\right)$$
$$= \max\left[\sum_{a,b,c...} l_a m_b n_c G_{fs,a,b,c}\right] \tag{23}$$

We used here the orthogonality of the Pauli matrices and find that by describing any witness $\hat{G}_{fs}$ in the HS decomposition of Eq. (23), the parameter $\alpha_{fs}$ is given by the right side of Eq. (23). The parameters $l_a m_b n_c ...$ (with subscripts a, b, c denoting x, y, or z) are given by Eq. (18) for 3-qubits states (or similar expressions for $n > 3$ qubits).

An entanglement witness for full separability $EW^{fs}$ gives by using Eq. (16) the probability ${p_{crit}}^{fs}$ where for $p \geq {p_{crit}}^{fs}$ entangled states mixed with white noise are not fully-separable, but it does not give information if $p < {p_{crit}}^{fs}$. We find that the witness for full separability ${E_{wit}}^{fs}$ gives similar results to those of PT but might be better. We find that in order to confirm the optimal value for full-separability we need to add explicit calculations transforming the density matrix to the form of Eq. (2) which will give the explicit condition ${p_{\exp}}^{fs}$ *below* which (or equal to it ) the state is $fs$.

## 3. Bell operators for $\left|GHZ(3)\right\rangle$ , $\left|W(3)\right\rangle$ and $\left|Cl_4\right\rangle$ states

**A.** For $\left|GHZ(3)\right\rangle$ [18, 19] the best choice for the Bell operator seems to be the 3-qubits correlations part of $\rho$ in the HS decomposition [7]:

$$\begin{aligned}\hat{B}_{G3} &= \left(\sigma_x\right)_A\otimes\left(\sigma_x\right)_B\otimes\left(\sigma_x\right)_C - \left(\sigma_x\right)_A\otimes\left(\sigma_y\right)_B\otimes\left(\sigma_y\right)_C \\ &-\left(\sigma_y\right)_A\otimes\left(\sigma_x\right)_B\otimes\left(\sigma_y\right)_C - \left(\sigma_y\right)_A\otimes\left(\sigma_y\right)_B\otimes\left(\sigma_x\right)_C\end{aligned} . \tag{24}$$

The full density matrix of $\left|GHZ(3)\right\rangle$ in the HS decomposition is [7]:

$$\begin{aligned}
&8\rho_{G3}=(I)_A\otimes(I)_B\otimes(I)_C+\hat{B}_{G,3}+(I)_A\otimes(\sigma_z)_B\otimes(\sigma_z)_C\\
&(\sigma_z)_A\otimes(I)_B\otimes(\sigma_z)_C+(\sigma_z)_A\otimes(\sigma_z)_B\otimes(I)_C\\
&R_{033}=R_{303}=R_{033}=R_{111}=1\ ;\ R_{122}=R_{212}=R_{221}=-1
\end{aligned} \quad . \qquad (25)$$

Note that $\hat{B}_{G3}$ is identical to the Mermin-Bell operator [30] (but obtained here by a different approach) and that $\left|GHZ(3)\right\rangle$ is an eigenstate of $\hat{B}_{G,3}$ with eigenvalue 4. The quantum limit is 4 while the classical bound is 2 so that the ratio between the classical bound and the quantum limit is 1/2. Hence $\left|GHZ(3)\right\rangle$ mixed with white has no LHV model for p>1/2.

**B.** We choose the Bell operator for the state: $\left|W(3)\right\rangle=\left(1/\sqrt{3}\right)\left(\left|100\right\rangle+\left|010\right\rangle+\left|001\right\rangle\right)$ as a part of its HS decomposition. The HS decomposition of $\left|W(3)\right\rangle$ was given in our previous paper ([7] Eq. (5.3)). It has 19 non-vanishing terms:

$$\begin{aligned}
&R_{113}=R_{131}=R_{311}=R_{223}=R_{213}=R_{322}=R_{011}=R_{101}=R_{110}=R_{022}=R_{202}=R_{220}=\frac{2}{3};\\
&R_{033}=R_{303}=R_{330}=-\frac{1}{3};\quad R_{333}=-1;\quad R_{003}=R_{030}=R_{300}=\frac{1}{3}
\end{aligned} \quad . \qquad (26)$$

Here the subscripts 1, 2, 3 refer to products with Pauli matrices $\sigma_{x,}\sigma_{y,}\sigma_z$ ,respectively, and the subscript 0 refers to product with the 2x2 unit matrix. The HS decomposition is quite complicated but it helps us to choose Bell operators as a part of the HS decomposition. A simple Bell operator for $\left|W(3)\right\rangle$ is obtained by choosing some of the 3-qubits correlations in its HS decomposition [7], with a slight change of numerical coefficients, for example

$$\begin{aligned}
&\hat{B}_{W3}=(\sigma_z)_A\otimes(\sigma_x)_B\otimes(\sigma_x)_C+(\sigma_x)_A\otimes(\sigma_z)_B\otimes(\sigma_x)_C+(\sigma_x)_A\otimes(\sigma_x)_B\otimes(\sigma_z)_C\\
&-(\sigma_z)_A\otimes(\sigma_z)_B\otimes(\sigma_z)_C
\end{aligned} \quad . \qquad (27)$$

Here the quantum limit $\beta_{Qu}=\left\langle W(3)\middle|\hat{B}_{W3}\middle|W(3)\right\rangle=3$, while the classical bound $\beta_{cl}=2$ so we get $p_{crit}=\frac{\beta_{cl}}{\beta_{Qu}}=2/3$ . Hence $\left|W(3)\right\rangle$ mixed with white has no LHV model for p>2/3. This is not necessarily the optimal value.

**C.** The density matrix of $2|Cl_4\rangle = |0000\rangle + |1100\rangle + |0011\rangle - |1111\rangle$ in the HS decomposition can be written as

$$16\rho_{Cl4} = (I)_A \otimes (I)_B \otimes (I)_C \otimes (I)_D + \sum_{\substack{a,b,c,d=0\\(a.b,c,d)\neq(0,0,0,0)}}^{3} R_{a.b.c,d} (\sigma_a)_A \otimes (\sigma_b)_B \otimes (\sigma_c)_C \otimes (\sigma_d)_D . \quad (28)$$

Here we have 15 non-vanishing HS parameters:

$$\begin{aligned} &R_{1130} = R_{1103} = R_{3011} = R_{0311} = R_{3300} = R_{0033} = R_{3333} == R_{1212} = R_{1221} = R_{2112} = R_{2121} = 1 \\ &R_{2230} = R_{2203} = R_{3022} = R_{0322} = -1 \end{aligned} . \quad (29)$$

A judicious choice of the Bell operator consists of the following 8 terms from the HS decomposition:

$$\begin{aligned} \hat{B}_{Cl4} = &\left[ (\sigma_x)_A \otimes (\sigma_y)_B + (\sigma_y)_A \otimes (\sigma_x)_B \right] \otimes \left[ (\sigma_x)_C \otimes (\sigma_y)_D + (\sigma_y)_C \otimes (\sigma_x)_D \right] + \\ &\left[ (\sigma_x)_A \otimes (\sigma_x)_B - (\sigma_y)_A \otimes (\sigma_y)_B \right] \otimes \left[ (\sigma_z)_C \otimes (I)_D + (I)_C \otimes (\sigma_z)_D \right] \end{aligned} \quad (30)$$

Calculating $\beta_{Cl}$ for this Bell operator we note that the first squared brackets in the first and second lines cannot be nonzero simultaneously. Hence the classical bound is 4 while the quantum value is: $Tr(\rho \hat{B}_{Cl4}) = 8$. So the ratio between the classical bound and the quantum value is 1/2. By mixing the cluster state with white noise with probability $p$ we find that there is no LHV model for $CL_4$ for $p > 1/2$.

## 4. Partial transpose, witness and explicit full-separability of $|GHZ(3)\rangle$) and $|W(3)\rangle$ states

### A. GHZ(3)

We define the entanglement witness for $|GHZ(3)\rangle$ by

$$EW_{G3}^{fs} = \alpha_{G3}^{fs} (I)_A \otimes (I)_B \otimes (I)_C - \hat{G}_{fs}^{G3} \;; \hat{G}_{fs}^{G3} = \hat{B}_{G,3} + (\sigma_z)_A \otimes (\sigma_z)_B \otimes (I)_C \quad (31)$$

Here we used entanglement witness for full separability $EW^{fs}$ according to Eq. (11). The Bell operator $\hat{B}_{G,3}$ for $|GHZ(3)\rangle$ is given by Eq. (24). Note that we added the term $(\sigma_z)_A \otimes (\sigma_z)_B \otimes (I)_C$ , which appears in the HS decomposition, to $\hat{B}_{G3}$ in the definition of $EW_{fs}^{G3}$ (The addition of this term improves the use of $EW_{fs}^{G3}$ but spoils the use of the Bell operator for LHV. To apply Eq. (31) for calculating $p_{crit}{}^{fs}$ according to Eq. (16) we notice that the present choice of the witness operator $\hat{G}_{fs}{}^{G3}$ satisfies

$$Tr\left[\hat{G}_{fs}{}^{G3}\right]=0 \tag{32}$$

since it is composed of products of Pauli matrices whose trace vanishes. In the present case Eq. (16) is reduced to a simple form given by

$$p_{wit,cr}{}^{fs}=\frac{\alpha^{fs}}{Tr\left(\rho_{Ent}G_{fs}\right)} \quad . \tag{33}$$

To calculate the parameter $\alpha_{G3}{}^{fs}$ we insert the components of $\hat{G}_{G,3}{}^{fs}$ into Eq. (23) and then we get:

$$\begin{aligned}\alpha_{G3}{}^{fs} &= \max\left(l_x m_x n_x - l_x m_y n_y - l_y m_x n_y - l_y m_y n_x + l_z m_z\right) = \\ &\max\begin{bmatrix}\sin\theta_A \sin\theta_B \sin\theta_C[\cos\phi_A\left(\cos\phi_B\cos\phi_C - \sin\phi_b\sin\phi_C\right)_B \\ -\sin\phi_A\left(\cos\phi_B\sin\phi_c + \sin\phi_B\cos\phi_C\right)] + \cos\theta_A\cos\theta_B\end{bmatrix} = 1\end{aligned} \tag{34}$$

Here, we substituted the $l,m,n$ terms according to Eq. (18). It is easily shown that the expression in the big square brackets of Eq. (34) is bounded by a cosine function.

Using the density matrix of the GHZ state $\rho_{G,3}$ of Eq. (31) and $\hat{G}_{fs}{}^{G3}$ according to Eq. (31) we get

$$Tr(\hat{G}_{fs}{}^{G3}\rho_{G3})=5 \quad . \tag{35}$$

Here we used the simple orthonormal relations of the Pauli matrices. In conclusion we get:

$$p_{cr}=\frac{\alpha_{G3}{}^{fs}}{Tr\left[G_{fs}{}^{G3}\rho_{G3}\right]}=0.2 \quad . \tag{36}$$

Therefore for $p>0.2$ , $\rho_{WN,G3}$ is not fully separable.

In a previous article [8,31] we have shown that $p=0.2$ is a sufficient condition for full separability of $\left|GHZ(3)\right\rangle$ state mixed with white noise by describing its density matrix explicitly in the form of Eq. (2) (see also [32]). We find here that this condition is also necessary so that $p\leq 0.2$ is a sufficient and necessary condition for full separability (see also equivalent result in [33, 34]). It is interesting to note that by using $PT$ for $\left|GHZ(3)\right\rangle$ relative to qubit $A$ ($PTA$)

we find [7] that this state is not fully separable for $p > 0.2$ so that this result is in agreement with the results for $EW_{G3}{}^{fs}$.

**B. W(3)**

For analyzing the full separability for $|W(3)\rangle$ we define $G_{fs}{}^{W3}$ for 3-qubits W state by

$$\begin{aligned}\hat{G}_{fs}{}^{W3} &= (\sigma_z)_A \otimes (\sigma_x)_B \otimes (\sigma_x)_C + (\sigma_x)_A \otimes (\sigma_z)_B \otimes (\sigma_x)_C + (\sigma_x)_A \otimes (\sigma_x)_B \otimes (\sigma_z)_C \\ &- (\sigma_z)_A \otimes (\sigma_z)_B \otimes (\sigma_z)_C + (\sigma_z)_A \otimes (\sigma_y)_B \otimes (\sigma_y)_C + \\ &(\sigma_y)_A \otimes (\sigma_z)_B \otimes (\sigma_y)_C + (\sigma_y)_A \otimes (\sigma_y)_B \otimes (\sigma_z)\end{aligned} \tag{37}$$

We added here 3 products to the Bell operator which also appear in the HS decomposition of $|W(3)\rangle$. Such additional products improve the expression for $EW_{w3}{}^{fs}$ (not for Bell operator).

Using Eq. (23) for the present case we get after rearranging terms:

$$\begin{aligned}\alpha_{w3}{}^{fs} &= \max\left[l_z m_x n_x + l_x m_z n_x + l_x m_x n_z + l_z m_y n_y + l_y m_z n_y + l_y m_y n_z - l_z m_z n_z\right] \\ &= \max\left\{\begin{aligned}&\begin{bmatrix}\sin\theta_A cos\theta_B \sin\theta_C \cos\phi_A \cos\phi_C + \sin\theta_A \sin\theta_B \cos\theta_C \cos\phi_A \cos\phi_B \\ \sin\theta_A cos\theta_B \sin\theta_C \sin\phi_A \sin\phi_C + \sin\theta_A \sin\theta_B \cos\theta_C \sin\phi_A \sin\phi_B \\ \cos\theta_A \sin\theta_B \sin\theta_C \cos\phi_B \cos\phi_C + \cos\theta_A \sin\theta_B \sin\theta_C \sin\phi_B \sin\phi_C\end{bmatrix} \\ &-cos\theta_A cos\theta_B \cos\theta_C\end{aligned}\right\} = 1\end{aligned} \quad . \tag{38}$$

Using again Eq. (16) for $|W(3)\rangle$ mixed with white noise we get

$$p_{crit}{}^{W3} = \frac{\alpha_{w3}{}^{fs}}{\langle W3|G_{fs}{}^{w3}|W3\rangle} = \frac{1}{5} = 0.2 \quad . \tag{39}$$

The value 5 follows from the explicit HS decomposition of the state $|W3\rangle$ [7], where the expectation value for $-(\sigma_z)_A \otimes (\sigma_z)_B \otimes (\sigma_z)_C$ is 1 and the other 6 expectation values of $\sigma's$ products in the denominator of Eq. (36) add to $6 \times \frac{2}{3} = 4$ (total number 5). Since the value $p_{crit} = 0.2$ is both sufficient (by explicit calculation) and necessary (by the full separability witness) then this value is optimal. It is interesting to note that by using the PT transformation it was shown [7] thar the critical value for full separability $p = 0.209589$ which is a little above the optimal value (see also [35]).

## 5. **The use of $PT$ , full-separability entanglement witness and explicit full separability construction for the cluster state mixed with white noise**

The density matrix of $2\left|Cl_4\right\rangle = \left|0000\right\rangle + \left|1100\right\rangle + \left|0011\right\rangle - \left|1111\right\rangle$ was given in Eqs. (28-29).

The use of 3 methods for full separability/non-full-separability of this state is demonstrated as follows:

### A. The use of PTA for $\left|Cl_4\right\rangle$ mixed with white noise:

We note that any $\rho$ of n-qubit state may be written as [7]:

$$2^n \rho = (I)^n + G + S \tag{40}$$

where under a given $PT$ (say with respect to A) $(I)^n + G$ is unchanged, and $S \to -S$. Here $(I)^n$ represents in a short notation the product $(I)_A \otimes (I)_B \otimes (I)_C \cdots$. Then under $PTA$ $2^n \rho$ is changed to

$$2^n \rho(PTA) = (I)^n + G - S \equiv 2^n \rho - 2S \quad . \tag{41}$$

By using $PTA$ of $\left|Cl_4\right\rangle$ only the HS terms that include $(\sigma_y)_A$ in the HS decomposition invert their sign so that they are included in $S$ :

$$\begin{aligned} S^{CL4} &= (\sigma_y)_A \otimes (\sigma_x)_B \otimes (\sigma_x)_C \otimes (\sigma_y)_D + (\sigma_y)_A \otimes (\sigma_x)_B \otimes (\sigma_y)_C \otimes (\sigma_x)_D \\ &- (\sigma_y)_A \otimes (\sigma_y)_B \otimes (\sigma_z)_C \otimes (I)_D - (\sigma_y)_A \otimes (\sigma_y)_B \otimes (I)_C \otimes (\sigma_z)_D . \\ 16\rho_{Cl4}{}^{PTA} &= (I)^n + G - S \equiv 16\rho_{Cl4} - 2S^{Cl4} \end{aligned} \quad . \tag{42}$$

Using $PTA$ for the $\left|Cl_4\right\rangle$ state with white noise then Eq. (6) is transformed to

$$\begin{aligned} 16\rho_{WN,Cl4}{}^{PTA} &= (1-p)(I)_A \otimes (I)_B \otimes (I)_C \cdots + 16p\rho_{Cl4}{}^{PTA} \ ; \\ 16\rho_{Cl4}{}^{PTA} &= 16\rho_{Cl4} - 2S^{Cl4} \end{aligned} \quad . \tag{43}$$

The eigenvalues of $16\rho_{WN,Cl4}{}^{PTA}$ were calculated and are given by

$$\begin{aligned} &\lambda_1 = 1-9p \ ; \ \lambda_2 = 1-7.4155p \ ; \ \lambda_3 = 1+3.4397p \ ; \ \lambda_4 = 1+7p \ ; \lambda_5 = 1+16.9755p \ ; \\ &\lambda_6 = \lambda_7 = \lambda_8 = \lambda_9 = \lambda_{10} = \lambda_{11} = \lambda_{12} = \lambda_{13} = \lambda_{14} = \lambda_{15} = \lambda_{16} = 1-p \end{aligned} . \tag{44}$$

From the lowest eigenvalue $\lambda_1 = 1-9p$ we find that for $p > \frac{1}{9}$ PTA of $\left|Cl_4\right\rangle$ gives negative eigenvalue so that under this condition it cannot be fully separable.

**B. The use of non-full separability entanglement witness for $|Cl_4\rangle$ state mixed with white noise:**

The Bell operator for the state $|Cl_4\rangle$ was given by Eq. (30). It includes 8 terms out of the 15 terms in the HS decomposition of $|Cl_4\rangle$.

The entanglement witness for full separability is chosen to be:

$$\begin{aligned}\hat{G}_{fs}^{Cl4} &= \left[(\sigma_x)_A \otimes (\sigma_y)_B + (\sigma_y)_A \otimes (\sigma_x)_B\right] \otimes \left[(\sigma_x)_C \otimes (\sigma_y)_D + (\sigma_y)_C \otimes (\sigma_x)_D\right] \\ &+ \left[(\sigma_x)_A \otimes (\sigma_x)_B - (\sigma_y)_A \otimes (\sigma_y)_B\right] \otimes \left[(\sigma_z)_C \otimes (I)_D\right] + \left[(I)_C \otimes (\sigma_z)_D\right] \otimes \\ &\left[(\sigma_x)_C \otimes (\sigma_x)_D - (\sigma_y)_C \otimes (\sigma_y)_D\right] + (\sigma_z)_A \otimes (\sigma_z)_B \otimes (\sigma_z)_C \otimes (\sigma_Z)_D\end{aligned} \quad . \quad (45)$$

We added here two products to the Bell operator which also appear in the HS decomposition of $|Cl_4\rangle$. Such additional products improve the expression for $EW^{fs}{}_{|Cl_4\rangle}$ (not for Bell operator). Using Eq. (23) for the present case we get after rearranging terms

$$\alpha_{w3}{}^{fs} = \max \begin{bmatrix} l_x m_y n_x o_y + l_x m_y n_y o_x + l_y m_x n_x o_y + l_y m_x n_y o_x \\ + l_x m_x n_z I + I l_z m_x n_x + l_z m_z n_z o_z - l_y m_y n_z I - I m_z n_y o_y \end{bmatrix} \quad . \quad (46)$$

We substitute the trigonometric functions for the $l,m,n,o$ terms according to Eq. (18) (including those for the $o$ terms) and after a straightforward calculation we get that the maximal value for the expression in the square brackets of Eq. (46) is 1. So we obtained:

$$\alpha_{Cl4}{}^{fs} = 1 \quad . \quad (47)$$

For $|Cl_4\rangle$ mixed with white noise we get according to Eq. (33)

$$p_{crit}{}^{Cl4} = \frac{\alpha_{fs}{}^{Cl4}}{\langle Cl_4 | \hat{G}_{fs}{}^{Cl4} | Cl_4 \rangle} = \frac{1}{9} \quad . \quad (48)$$

Here we substituted the value $\alpha_{fs}{}^{Cl4} = 1$ from Eq. (47). The density matrix of the state $|Cl_4\rangle$ is given by Eqs. (28, 29) and $\hat{G}_{fs}{}^{Cl4}$ is given by Eq. (45). By using full separability entanglement witness we find here that the $|Cl_4\rangle$ state mixed with white noise cannot be fully separable for $p > \frac{1}{9}$. This result is equivalent to that obtained by $PT$.

**C.** We describe In Appendix A explicit construction for full separability of $\left|Cl_4\right\rangle$ mixed with white noise. We find that explicit construction for full separability of $\left|Cl_4\right\rangle$ mixed with white noise can be obtained for: $p \le 1/9$. On the other hand it followed from the $PT$ analysis and from the non-full separability entanglement witness analysis that for $p > 1/9$ the $Cl_4$ state mixed with white noise is not fully separable. Hence the critical value for EW is $p_{crit} = 1/9$. Above it, it is not fully separable, below it, it is fully separable.

## 6. Entanglement witness and explicit construction for bi-separability of 3-qubits states

A certain partition of a pure 3-qubits bi-separable state is given by Eq. (3) as $\phi^{bs} = \phi_{AB} \otimes \phi_C$ where $\phi_{AB}$ is a pure state of qubits $A$ and $B$, and $\phi_C$ is a pure state of qubit $C$ given by Eq. (4). $\phi_{AB}$ can be written as

$$\phi_{AB} = \sum_{i=1}^{4} a_i \phi_{AB}{}^{(i)} \tag{49}$$

where

$$\begin{array}{l}
\phi_{AB}{}^{(1)} = \begin{pmatrix} 1 \\ 0 \end{pmatrix}_A \otimes \begin{pmatrix} 1 \\ 0 \end{pmatrix}_B = \begin{pmatrix} 1 \\ 0 \\ 0 \\ 0 \end{pmatrix} \quad ; \quad \phi_{AB}{}^{(2)} = \begin{pmatrix} 1 \\ 0 \end{pmatrix}_A \otimes \begin{pmatrix} 0 \\ 1 \end{pmatrix}_B = \begin{pmatrix} 0 \\ 1 \\ 0 \\ 0 \end{pmatrix} \\
\phi_{AB}{}^{(3)} = \begin{pmatrix} 0 \\ 1 \end{pmatrix}_A \otimes \begin{pmatrix} 1 \\ 0 \end{pmatrix}_B = \begin{pmatrix} 0 \\ 0 \\ 1 \\ 0 \end{pmatrix} \quad ; \quad \phi_{AB}{}^{(4)} = \begin{pmatrix} 0 \\ 1 \end{pmatrix}_A \otimes \begin{pmatrix} 0 \\ 1 \end{pmatrix}_B = \begin{pmatrix} 0 \\ 0 \\ 0 \\ 1 \end{pmatrix}
\end{array} . \tag{50}$$

The parameters $a_i$ are the components of the 4-dimensional vector satisfying the relation $\sum_{i=1}^{4} |a_i|^2 = 1$. A general representation of these parameters can be given as:

$$a_1 = \cos\theta_1 \;\; ; \;\; a_2 = e^{i\varphi_2}\sin\theta_1\cos\theta_2 \;\; ; a_3 = e^{i\varphi_3}\sin\theta_1\sin\theta_2\cos\theta_3$$
$$a_4 = e^{i\varphi_4}\sin\theta_1\sin\theta_2\sin\theta_3 \quad . \tag{51}$$

In analogy to Eq. (11) for a full-separability witness, we define here the bi-separability witness ${E_{wit}}^{bs}$ for genuine entanglement by

$${E_{wit}}^{bs} = \alpha^{bs}\left(I\right)^n - G_{bs} \quad . \tag{52}$$

where for n-qubits $I$ and $G_{bs}$ are the unit and a certain Hermitian operator, respectively, with dimension $2^n$, and $\alpha_{bs}$ is a certain parameter, The superscript and subscripts $bs$ refer to bi-separable states.

In a similar way to Eqs. (8-9) if for any $\phi^{bs}$

$$Tr\left({E_{wit}}^{bs}\left|\phi^{bs}\right\rangle\left\langle\phi^{bs}\right|\right) \geq 0 \quad \left(\Rightarrow \alpha_{bs} = \max Tr\left(\left|\phi^{bs}\right\rangle\left\langle\phi^{bs}\right| G_{bs}\right)\right) \tag{53}$$

and

$$Tr({E_{wit}}^{bs}\rho) < 0. \tag{54}$$

*then* $\rho$ *is genuinely entangled*

By inserting $\rho_{WN}$ of Eq. (6) (representing entangled state $\rho_{Ent}$ mixed with white noise) and ${E_{wit}}^{bs}$ of Eq. (52) into Eq. (54) we get for n-qubits states mixed with white noise the condition for genuine entanglement:

$$\max Tr\left[\left(\alpha^{bs}\left(I\right)^n - G_{bs}\right)\left(\left(\frac{1-p}{2^3}\right)\left(I\right)^n + p\rho_{Ent}\right)\right] < 0 \quad . \tag{55}$$

Eq, (55) is analogous to Eq. (13) where we exchange $\alpha^{fs}$ and $G_{fs}$ to $\alpha^{bs}$ and $G_{bs}$, respectively. Following analogous procedure to that in which Eq. (16) was derived for $fs$ we get here for $bs$ states the limiting critical value ${p_{cr}}^{bs}$ above which the system is genuinely entangled

$${p_{cr}}^{bs} = \frac{\alpha^{bs} - \frac{TrG_{bs}}{2^n}}{Tr\left(\rho_{Ent}G_{bs}\right) - \frac{TrG_{bs}}{2^n}} \quad . \tag{56}$$

Under the condition that $G_{bs}$ is trace-less Eq. (56) reduces to

$$p_{cr}{}^{bs} = \frac{\alpha_{bs}}{Tr\left(\rho_{Ent} G_{bs}\right)} \quad . \tag{57}$$

It is a quite difficult task to find the witness $G_{bs}$ which by calculating $\alpha^{bs}$ and using Eq. (56), or Eq. (57) (for trace-less witness operator), we will get the optimal critical value $p_{cr}{}^{bs}$.

# 7. Entanglement witness and explicit construction for bi-separability of $|GHZ,3\rangle$ state

Using the Bell states of the subsystems $AB, BC, AC$ we define three explicitly bi-separable pure density matrices. The density matrix of an entangled state with probability $p$ mixed with white noise with probability $1-p$ is

$$\rho_{WN}{}^{G3} = \frac{1-p}{2^n}\left(I\right)_A \otimes \left(I\right)_B \otimes \left(I\right)_C + p\rho^{G3} = \frac{1-p}{2^n}\left(I\right)_A \otimes \left(I\right)_B \otimes \left(I\right)_C$$

$$p\left[\frac{1}{2^n}\left(I\right)_A \otimes \left(I\right)_B \otimes \left(I\right)_C + \sum_{\substack{a,b,c\cdots=0 \\ (a,b,c\cdots)\neq(0,0,0\cdots)}}^{3} R_{a,b,c}{}^{G3}\left(\sigma_a\right)_A \otimes \left(\sigma_b\right)_B \otimes \left(\sigma_c\right)_C\right] \quad ; \sigma_0 = I \quad . \tag{58}$$

Here the subscript $WN$ denotes mixing with white noise. The parameters $R_{a,b,c}{}^{G3}$ are obtained from the HS decomposition for the density matrix $\rho_{Ent}$. This equation can be shortened into the form:

$$8\rho^{G3}{}_{WN} = \left(I\right)_A \otimes \left(I\right)_B \otimes \left(I\right)_C + pR$$

$$R = \left[\sum_{\substack{a,b,c\cdots=0 \\ (a,b,c\cdots)\neq(0,0,0\cdots)}}^{3} R_{a,b,c}{}^{G3}\left(\sigma_a\right)_A \otimes \left(\sigma_b\right)_B \otimes \left(\sigma_c\right)_C\right] \quad . \tag{59-a}$$

We find that the terms in the square brackets of Eq. (59-a) can be given by sum of bi-separable density matrices minus $-g\left(I\right)_A \otimes \left(I\right)_B \otimes \left(I\right)_C$ where g is a certain coefficient. The total coefficient of $\left(I\right)_A \otimes \left(I\right)_B \otimes \left(I\right)_C$ on the right side of Eq. (59-a) becomes $1-pg$ and since this coefficient should be non-negative we find that for $p\left(bs\right) \leq 1/g$ the state is bi-separable (not genuinely entangled).

In Section (6) we used the special partition of the bi-separable state given by $\phi^{bs} = \phi_{AB} \otimes \phi_C$. In principle, for using the entanglement witness we should consider also other partitions such as $\phi^{bs} = \phi_{BC} \otimes \phi_A$ and/or $\phi^{bs} = \phi_{AC} \otimes \phi_B$. However, if the entanglement witness for bi-separability is unchanged by any permutation between the qubits $A, B, C$ then it is enough to take into account only one partition i.e. we consider only the partition: $\phi^{bs} = \phi_{AB} \otimes \phi_C$ (see Eqs. [3-4]).

a) **Witness for $|GHZ, 3\rangle$**

A possible choice for the witness $E_{wit}{}^{bs}$ for the state $|GHZ, 3\rangle$ is given by:

$$G_{bs}(GHZ,3) = 8\rho_{GHZ,3} - (I)_A \otimes (I)_B \otimes (I)_C \quad , \tag{60}$$

which is a trace-less operator, where $\rho_{GHZ,3}$ is the density matrix of the $|GHZ, 3\rangle$ state. We notice according to Eqs. (3-4. 51-53):

$$|\phi^{bs}\rangle = \begin{pmatrix} a_1 \cos\theta_C / 2 \\ e^{i\phi_C} a_1 \sin\theta_C / 2 \\ a_2 \cos\theta_C / 2 \\ e^{i\phi_C} a_2 \sin\theta_C / 2 \\ a_3 \cos\theta_C / 2 \\ e^{i\phi_C} a_3 \sin\theta_C / 2 \\ a_4 \cos\theta_C / 2 \\ e^{i\phi_C} a_4 \sin\theta_C / 2 \end{pmatrix} \quad . \tag{61}$$

According to Eqs. (53, 58) we get:

$$\alpha_{bs} = \max\langle\phi_{bs}|\hat{G}|\phi_{bs}\rangle = \max\left[8\langle\phi_{bs}|\rho_{GHZ,3}|\phi_{bs}\rangle - 1\right] = \max\left[8\,|\langle GHZ,3|\phi_{bs}\rangle^2| - 1\right] \quad . \tag{62}$$

Using Eqs. (61,62), we get:

$$\begin{aligned} &\sqrt{2}\left|\langle GHZ(3)|\phi_{bs}\rangle\right| \le \left|a_1 \cos\left(\frac{\theta_C}{2}\right) + a_4 e^{i\phi_C} \sin\left(\frac{\theta_C}{2}\right)\right| \\ &\le \left|\cos\theta_1 \cos\left(\frac{\theta_C}{2}\right)\right| + \left|\sin\theta_1 \sin\left(\frac{\theta_C}{2}\right)\right| \le 1 \end{aligned} \quad . \tag{63}$$

Hence

$$\alpha^{bs} == 4-1=3 \quad . \tag{64}$$ .

Using Eq, (57) for $p_{cr}{}^{bs}$ we get:

$$p_{cr}{}^{bs,G3} = \frac{\alpha_{bs}{}^{G3}}{Tr\left[\hat{G}_{GHZ,3}\rho_{Ent}\right]} = \frac{3}{7} \quad . \tag{65}$$

where $G3$ is a short notation for $GHZ(3)$. This result is equivalent to that obtained previously by other authors (e.g. [20]).

**b)** In order to show that this result is optimal we show in Appendix B that an explicit construction of bi-separability of $|GHZ,3\rangle$ state mixed with white noise gives the same result i.e. that $\rho^{G3}{}_{WN}$ is separable for $p \leq 3/7$.

## 8. Entanglement witness and explicit construction for bi-separability of $|W(3)\rangle$

**a) Analysis of a witness for** $|w(3)\rangle$

The $|w(3)\rangle$ state is $\sqrt{3}|w(3)\rangle = |100\rangle + |010\rangle + |001\rangle$ and its density matrix is given as $\rho_{W3} = |w\rangle\langle w|$. A simple choice for the witness for the $|w(3)\rangle$ state is given by $G_{wit}{}^{W3} = \rho_{W3}$ i.e.

$$G_{wit}{}^{W3} = \frac{1}{3}\left(|100\rangle + |010\rangle + |001\rangle\right)\left(\langle 100| + \langle 010| + \langle 001|\right) \quad . \tag{66}$$

Using Eq. (53) we get

$$\alpha_{bs}{}^{w3} = \max\left\langle \phi^{bs}\right| G_{wit}{}^{w3} \left|\phi^{bs}\right\rangle = \max\left|\left\langle \phi^{bs} | w(3)\right\rangle\right|^2 \quad . \tag{67}$$

By straightforward calculations of Eq. (67) we get:

$$\sqrt{3}\left\langle W(3)|\phi^{bs}\right\rangle = a_1 e^{i\phi_C}\sin\left(\frac{\theta_C}{2}\right) + (a_2 + a_3)\cos\left(\frac{\theta_C}{2}\right) \tag{68}$$

where we substitute $a_1, a_2, a_3$ according to Eq. (51). Then we get

$$\max\left|\left\langle W(3)|\phi^{bs}\right\rangle\right| = \max\left\{\left|\cos\theta_1 \sin\frac{\theta_C}{2}\right| + |\left(\sin\theta_1\cos\theta_2 + \sin\theta_1\sin\theta_2\right)\cos\frac{\theta_C}{2}|\right\} = \sqrt{2} \quad . \tag{69}$$

(We satisfy this result by: $\sin\theta_1 = 1;\ \cos\theta_3 = \cos\frac{\theta_C}{2} = 1;\ \sin\theta_2 = 1/\sqrt{2} = \cos\theta_2$) . Hence

$\alpha_{bs}{}^{w3} = 2/3$. By using Eq. (56) we get:

$$p_{cr,bs}{}^{w3} = \frac{2/3-1/8}{1-1/8} = \frac{13}{21} \quad . \tag{70}$$

This result was obtained already by other authors (e.g. [20]). By changing $G_{wit}{}^{W3}$ to: $G_{wit}{}^{W3} = \frac{2}{3}|111\rangle\langle 111| + \rho_{w3}$ it was found [34] that $\alpha_{bs}{}^{w3} = 2/3$ is not changed but $TrG_{wit}{}^{W3} = 1 + 2/3$ so $p_{cr,bs}{}^{w3} = 11/19$. The value $p_{cr,bs}{}^{w3} = 11/19$ was reduced to around 0.529 by using other methods [28]. One should note that $p$ defined by us is equal to $1-p$ defined by other authors [20, 28, 34].

**b) Explicit construction for bi-separability of $|W(3)\rangle$ mixed with white noise:**

We find for this case that the terms in the square brackets of Eq. (59-b) below for $\rho_{WN}{}^{W3}$ can be given in a similar way for $\rho_{WN}{}^{G3}$, given as

$$8\rho^{W3}{}_{WN} = (I)_A \otimes (I)_B \otimes (I)_C + pR$$
$$R = \left[ \sum_{\substack{a,b,c\cdots=0 \\ (a,b,c\cdots)\neq(0,0,0\cdots)}}^{3} R_{a,b,c}{}^{W3} (\sigma_a)_A \otimes (\sigma_b)_B \otimes (\sigma_c)_C \right] \quad . \tag{59-b}$$

A part of $\sum_{\substack{a,b,c\cdots=0 \\ (a,b,c\cdots)\neq(0,0,0\cdots)}}^{3} R_{a,b,c}{}^{W3} (\sigma_a)_A \otimes (\sigma_b)_B \otimes (\sigma_c)_C$ is given by summing 3 bi-separable un-normalized density matrices ($\rho_1, \rho_2, \rho_3$) minus the term $\frac{2}{3}(I)_A \otimes (I)_B \otimes (I)_C$ from each of them. The density matrices $\rho_1, \rho_2, \rho_3$ are given by products of un-normalized density matrices of Bell states with the density matrix of the up state of the third qubit inserted after the Bell state, before

it or in the middle:

$$8\rho_1 = (\frac{2}{3})[(I)_A \otimes (I)_B + (\sigma_x)_A \otimes (\sigma_x)_B + (\sigma_y)_A \otimes (\sigma_y)_B - (\sigma_z)_A \otimes (\sigma_z)_B] \otimes [(I) + (\sigma_z)]_C$$

$$8\rho_2 = (\frac{2}{3})[(I) + (\sigma_z)]_A \otimes [(I)_B \otimes (I)_C + (\sigma_x)_B \otimes (\sigma_x)_C$$
$$+(\sigma_y)_B \otimes (\sigma_y)_C - (\sigma_z)_B \otimes (\sigma_z)_C] \tag{71}$$
$$8\rho_3 = (2/3)[(I)_A \otimes (I+\sigma_z)_B \otimes (I)_C + (\sigma_x)_A \otimes (I+\sigma_z)_B \otimes (\sigma_x)_C$$
$$+(\sigma_y)_A \otimes (I+\sigma_z)_B \otimes (\sigma_y)_C - (\sigma_z)_A \otimes (I+\sigma_z)_B \otimes (\sigma_z)_C]$$

By expanding the density matrices: $8\rho_1, 8\rho_2$, and $8\rho_3$ we get:

$$8\rho_1 + 8\rho_2 + 8\rho_3 - 2(I)_A \otimes (I)_B \otimes (I)_C = \sum_{\substack{a,b,c=0 \\ (a,b,c)\neq(0,0,0\cdots)}}^{3} R_{a,b,c}(\sigma_a)_A \otimes (\sigma_b)_B \otimes (\sigma_c)_C \tag{72}$$

where the parameters $R_{a,b,c}$ for $8\rho_{W3}$ are given by

$$\begin{aligned} &R_{113} = R_{131} = R_{311} = R_{223} = R_{232} = R_{322} \\ &= R_{011} = R_{101} = R_{110} = R_{022} = R_{202} = R_{220} = \frac{2}{3}; \\ &R_{033} = R_{303} = R_{330} = -\frac{2}{3}; \quad R_{333} = -2; \\ &R_{003} = R_{030} = R_{300} = \frac{2}{3} \end{aligned} \quad . \tag{73}$$

By comparing these values, with the $R_{a,b,c}$ parameters of the HS decomposition of Eq. (26) we find that the only difference is that we get here for 3 $R_{a,b,c}$ parameters -2/3 instead of -1/3, for 3 $R_{a,b,c}$ we get 2/3 instead of 1/3 and for $R_{333}$ we get coefficient -2 instead of -1.

To correct we add to $8\rho_1 + 8\rho_2 + 8\rho_3 - 2(I)_A \otimes (I)_B \otimes (I)_C$ the expansion of

$$8\rho_4 - (1/3)(I)_A \otimes (I)_B \otimes (I)_C \tag{74}$$

and

$$8\rho_5 - \frac{4}{3}(I)_A \otimes (I)_B \otimes (I)_C \tag{75}$$

where $8\rho_4$ and $8\rho_5$ are fully separable un-normalized density matrices given as

$$8\rho_4 = (1/3)\left[(I)_A \otimes (I)_B \otimes (I)_C + (I-\sigma_z)_A \otimes (I-\sigma_z)_B \otimes (I-\sigma_z)_C\right] \quad , \tag{76}$$

$$8\rho_5 = (4/3)[(I)_A \otimes (I)_B \otimes (I)_C + (\sigma_z)_A \otimes (\sigma_z)_B \otimes (\sigma_z)_C] =$$
$$(1/3)[(I+\sigma_z)_A \otimes (I+\sigma_z)_B \otimes (I+\sigma_z)_C + (I+\sigma_z)_A \otimes (I-\sigma_z)_B \otimes (I-\sigma_z)_B + \quad . \qquad (77)$$
$$(I-\sigma_z)_A \otimes (I-\sigma_z)_B \otimes (I+\sigma_z)_C + (I-\sigma_z)_A \otimes (I+\sigma_z)_B \otimes (I-\sigma_z)_B]$$

We insert in Eq, (59-b) defined above

$$R = 8(\rho_1 + \rho_2 + \rho_3 + \rho_4 + \rho_{\backslash 5}) - 3\frac{2}{3}(I)_A \otimes (I)_B \otimes (I)_C \quad . \qquad (78)$$

where $R$ gives all 19 HS parameters. Then we get by Eq. (59-b):

$$8\rho^{w3}{}_{WN} = (I)_A \otimes (I)_B \otimes (I)_C + p\left[8(\rho_1 + \rho_2 + \rho_3 + \rho_4 + \rho_5) - 3\frac{2}{3}(I)_A \otimes (I)_B \otimes (I)_C\right] \quad . \qquad (79)$$

So we find that for $p \le 3/11$ $|W(3)\rangle$ mixed with white noise is explicitly bi-separable since only under this condition the coefficient of $(I)_A \otimes (I)_B \otimes (I)_C$ in Eq. (79) for $\rho_{WN}{}^{W3}$ will be non-negative. On the other hand it was found [28] that for $p > 0.529$ this mixed state is genuinely entangled. We find that there is a large region $0.2727 \le p \le 0.529$ where we don't know if $\rho^{W3}{}_{WN}$ is bi-separable or genuinely entangled.

## 9. Summary and conclusions

We used the HS decomposition to define Bell operators (negating LHV models) and to define entanglement witnesses (negating full/bi separability; negating bi-separability implies genuine entanglement).

We treated entangled density matrix $\rho_{Ent}$ of n-qubits which is mixed with white noise, so that $\rho_{Ent}$ is changed to $\rho_{WN}$ given by Eq. (6) where the subscript WN denotes the admixture of white noise with probability $1-p$, and $\rho_{Ent}$ with probability $p$. We studied the critical value $p_{cr}{}^{fs}$ (Eq. (16)) above which the state is not fully separable. The use of $PT$ gives also $p_{cr}$ value above which the system is not fully separable. The use of $EW$ might give better results (or at least equal) to those obtained by $PT$. We studied these problems for: $|GHZ(3)\rangle$, $|W(3)\rangle$, and $|Cl_4\rangle$ .

As the choice of $EW$ is usually not optimal, we add to the analysis of $EW$ explicit construction of a fully separable density matrix and find the critical value $p_{cr}$ below which the system is certainly fully separable. Under the condition that this value coincides with $p_{cr}$ obtained by the EW we conclude that this parameter gives optimal result for a change from full separability to non-full separability.

The use of EW for genuine entanglement is analyzed by EW given by ${E_{wit}}^{bs} = \alpha^{bs}(I)^n - G_{bs}$ where $G_{bs}$ is a Hermitian operator. Genuine entanglement satisfies the criterions given by Eqs. (53-54). The parameter $\alpha_{bs}$ is given as $\alpha_{bs} = \max Tr\left(\left|\phi^{bs}\right\rangle\left\langle\phi^{bs}\right|G_{bs}\right)$, where $\left|\phi^{bs}\right\rangle$ is any bi-separable state given in Eq. (61).

We studied by genuine entangled witness ${E_{wit}}^{bs}$ the critical value ${p_{cr}}^{bs}$ <u>above</u> which the entangled state mixed with white noise is genuinely entangled. As the choice ${E_{wit}}^{bs}$ might not be optimal we added to the analysis explicit construction of a bi- separable density matrix and find the critical ${p_{cr}}^{bs}$ value below which the system is bi-separable. We find that for $|GHZ,3\rangle$ mixed with white noise the two values coincide giving ${p_{cr}}^{bs} = \frac{3}{7}$ so we conclude that this parameter is optimal.

For **t**he density matrix of $|W3\rangle$ with probability $p$ mixed with white noise with probability $1-p$ our best explicit calculation gave that for $p_{crit} \le \frac{3}{11}$ this state is bi-separable. On the other hand by the best witness obtained by other authors for this state [28] it was shown that above ${p_{cr}}^{bs} = 0.529$ this state is genuinely entangled. In the region $\frac{3}{11} \le p_{cr} < 0.529$ it is not clear to us if this state is genuinely entangled or only bi-separable.

**Appendix A: Explicit construction for full separability of $|Cl_4\rangle$ mixed with white noise:**

Using Eq. (6) $|Cl_4\rangle$ mixed with white noise is given by

$$16{\rho_{WN}}^{Cl4} = (1-p)(I)_A \otimes (I)_B \otimes (I)_C \otimes (I)_D + 16p\rho_{Cl4} \qquad \text{(A1)}$$

where $16\rho_{Cl4}$ is given by Eqs. (28,29) By substituting Eqs. (28,29) into Eq. (A1) we get

$$16\rho_{WN}{}^{Cl4} = (I)_A \otimes (I)_B \otimes (I)_C \otimes (I)_D +$$
$$p \sum_{\substack{a.b,c,d=0 \\ (a,b,c,d)\neq(0,0,0,0)}}^{3} R_{a.b.c,d} (\sigma_a)_A \otimes (\sigma_b)_B \otimes (\sigma_c)_C \otimes (\sigma_d)_D \quad . \tag{A2}$$

We choose from Eq. (A2) three products of Pauli matrices with coefficients $R_{1130} = R_{1103} = R_{0033} = 1$ and transform the sum of these products so that it will be given by the fully separable form in the squared brackets [denoted by: $FSF,1-(I)_A \otimes (I)_B \otimes (I)_C \otimes (I)_D$)] so we get

$$\begin{aligned}
&FSF,1-(I)_A \otimes (I)_B \otimes (I)_C \otimes (I)_D = \\
&(\sigma_x)_A \otimes (\sigma_x)_B \otimes (\sigma_z)_C \otimes (I)_D + (\sigma_x)_A \otimes (\sigma_x)_B \otimes (I)_C \otimes (\sigma_z)_D + \\
&(I)_A \otimes (I)_B \otimes (\sigma_z)_C \otimes (\sigma_z)_D \quad . \\
&= \left[ \frac{1}{4} \left\{ \begin{matrix} (I+\sigma_x)_A \otimes (I-\sigma_x)_B \otimes (I-\sigma_z)_C \otimes (I-\sigma_z)_D + \\ (I-\sigma_x)_A \otimes (I+\sigma_x)_B \otimes (I-\sigma_z)_C \otimes (I-\sigma_z)_D \\ +(I+\sigma_x)_A \otimes (I+\sigma_x)_B \otimes (I+\sigma_z)_C \otimes (I+\sigma_z)_D + \\ (I-\sigma_x)_A \otimes (I-\sigma_x)_B \otimes (I+\sigma_z)_C \otimes (I+\sigma_z)_D \end{matrix} \right\} \right] \\
&-(I)_A \otimes (I)_B \otimes (I)_C \otimes (I)_D
\end{aligned} \tag{A3}$$

Similarly, choosing three other products of Pauli matrices with coefficients $R_{3011} = R_{0311} = R_{3300} = 1$ it is transformed into the form

$$\begin{aligned}
&FSF,2-(I)_A \otimes (I)_B \otimes (I)_C \otimes (I)_D = \\
&(\sigma_z)_A \otimes (I)_B \otimes (\sigma_x)_C \otimes (\sigma_x)_D + (I)_A \otimes (\sigma_z)_B \otimes (\sigma_x)_C \otimes (\sigma_x)_D + \quad . \\
&(\sigma_z)_A \otimes (\sigma_z)_B \otimes (I)_C \otimes (I)_D
\end{aligned} \tag{A4}$$

It is straightforward to substitute an explicit form of $FSF,2$ but for simplicity we will use only this short notation. After lengthy and tedious algebra we find that the additional terms of Eq. (A2) of the form $R_{a.b.c,d} (\sigma_a)_A \otimes (\sigma_b)_B \otimes (\sigma_c)_C \otimes (\sigma_d)$ can be decomposed as $FSF,3-(I)_A \otimes (I)_B \otimes (I)_C \otimes (I)_D ;...FSF,9-(I)_A \otimes (I)_B \otimes (I)_C \otimes (I)_D$ which are fully separable expressions similar to Eq. (A3). Adding all the results for $FSF,j \; (j=1,2,...,9)$ and substituting them in Eq. (A2) we get:

$$\begin{aligned}
&16\rho_{WN}{}^{Cl4} = (I)_A \otimes (I)_B \otimes (I)_C \otimes (I)_D + \\
&+p\left[FSF,1+FSF,2+FSF,3+...,FSF,9\right] - 9p(I)_A \otimes (I)_B \otimes (I)_C \otimes (I)_D \quad .
\end{aligned} \tag{A5}$$

We find that explicit construction for full separability of $|Cl_4\rangle$ state mixed with white noise can be obtained for $p \le 1/9$ , since under this condition the coefficient of $(I)_A \otimes (I)_B \otimes (I)_C \otimes (I)$ is non-negative number. On the other hand it followed from the $PT$ analysis and from the non-full separability entanglement witness analysis given in Section 5 that for $p > 1/9$ the $Cl_4$ state mixed with white noise is not fully separable so that $p = 1/9$ is the critical value.

**Appendix B: Explicit construction for bi-separabilty of $|GHZ,3\rangle$ state**

**T**he density matrix of $|GHZ,3\rangle$ entangled state with probability $p$ mixed with white noise with probability $1-p$ is

$$\rho_{WN}{}^{G3} = \frac{1-p}{2^n} (I)_A \otimes (I)_B \otimes (I)_C + p\rho^{G3} = \frac{1-p}{2^n} (I)_A \otimes (I)_B \otimes (I) +$$
$$p\left[\frac{1}{2^n}(I)_A \otimes (I)_B \otimes (I)_C + \frac{1}{2^n} \sum_{\substack{a,b,c\cdots=0 \\ (a,b,c\cdots)\neq(0,0,0\cdots)}}^{3} R_{a,b,c}{}^{G3} (\sigma_a)_A \otimes (\sigma_b)_B \otimes (\sigma_c)_C\right] \quad ; \sigma_0 = I \ . \quad \text{(B1)}$$

Here the subscript $WN$ denotes mixing with white noise. The parameters $R_{a,b,c}{}^{G3}$ are obtained from the HS decomposition for the density matrix $\rho^{G3}$. This equation can be shortened as:

$$8\rho_{WN}{}^{G3} = (I)_A \otimes (I)_B \otimes (I)_C + pR$$
$$R = \left[\sum_{\substack{a,b,c\cdots=0 \\ (a,b,c\cdots)\neq(0,0,0\cdots)}}^{3} R_{a,b,c}{}^{G3} (\sigma_a)_A \otimes (\sigma_b)_B \otimes (\sigma_c)_C\right] \quad . \quad \text{(B2)}$$

We express $R$ of Eq. (B2) as the sum of 4 terms i.e.

$$R = 1/3(R_1 + R_2 + R_3 + R_4) \quad \text{(B3)}$$

where each $R_i (i = 1,2,3,4)$ is given by the sum of bi-separable density matrices minus $g(I)_A \otimes (I)_B \otimes (I)_C$ where $g$ is a certain coefficient.

Using Bell states of the subsystems $AB, BC, AC$ we define bi-sparable density matrices:

$$
\begin{aligned}
&8\rho_{A|BC} = \\
&\frac{1}{2}(I+\sigma_x)_A \otimes \left[(I)_B \otimes (I)_C + (\sigma_x)_B \otimes (\sigma_x)_C - (\sigma_y)_B \otimes (\sigma_y)_C + (\sigma_z)_B \otimes (\sigma_z)_C\right] \\
&+\frac{1}{2}(I-\sigma_x)_A \otimes \left[(I)_B \otimes (I)_C - (\sigma_x)_B \otimes (\sigma_x)_C + (\sigma_y)_B \otimes (\sigma_y)_C + (\sigma_z)_B \otimes (\sigma_z)_C\right] \\
&+\frac{1}{2}(I+\sigma_y)_A \otimes \left[(I)_B \otimes (I)_C - (\sigma_x)_B \otimes (\sigma_y)_C - (\sigma_y)_B \otimes (\sigma_x)_C + (\sigma_z)_B \otimes (\sigma_z)_C\right] \\
&+\frac{1}{2}(I-\sigma_y)_A \otimes \left[(I)_B \otimes (I)_C + (\sigma_x)_B \otimes (\sigma_y)_C + (\sigma_y)_B \otimes (\sigma_x)_C + (\sigma_z)_B \otimes (\sigma_z)_C\right]
\end{aligned}
\quad ; \quad \text{(B4)}
$$

$$
R_1 = 8\rho_{A|BC} - 2(I)_A \otimes (I)_B \otimes (I)_C \quad . \qquad \text{(B5)}
$$

$$
\begin{aligned}
&8\rho_{AB|C} = \\
&\frac{1}{2}\left[(I)_A \otimes (I)_B + (\sigma_x)_A \otimes (\sigma_x)_B - (\sigma_y)_A \otimes (\sigma_y)_B + (\sigma_z)_A \otimes (\sigma_z)_B\right] \otimes (I+\sigma_x)_C \\
&+\frac{1}{2}\left[(I)_A \otimes (I)_B - (\sigma_x)_A \otimes (\sigma_x)_B + (\sigma_y)_A \otimes (\sigma_y)_B + (\sigma_z)_A \otimes (\sigma_z)_B\right] \otimes (I-\sigma_x)_C \\
&+\frac{1}{2}\left[(I)_A \otimes (I)_B - (\sigma_x)_A \otimes (\sigma_y)_B - (\sigma_y)_A \otimes (\sigma_x)_B + (\sigma_z)_A \otimes (\sigma_z)_B\right] \otimes (I+\sigma_y)_C \\
&+\frac{1}{2}\left[(I)_A \otimes (I)_B + (\sigma_x)_A \otimes (\sigma_y)_B + (\sigma_y)_A \otimes (\sigma_x)_B + (\sigma_z)_A \otimes (\sigma_z)_B\right] \otimes (I-\sigma_y)_C
\end{aligned}
\quad ; \quad \text{(B6)}
$$

$$
R_2 = 8\rho_{ABC} - 2(I)_A \otimes (I)_B \otimes (I)_C \quad . \qquad \text{(B7)}
$$

$$
\begin{aligned}
&8\rho_{A|B|C} = \\
&\frac{1}{2}\begin{bmatrix} (I)_A \otimes (I+\sigma_x)_B \otimes (I)_C + (\sigma_x)_A \otimes (I+\sigma_x)_B \otimes (\sigma_x)_C \\ -(\sigma_y)_A \otimes (I+\sigma_x)_B \otimes (\sigma_y)_C + (\sigma_z)_A \otimes (I+\sigma_x)_B \otimes (\sigma_z)_C \end{bmatrix} \\
&+\frac{1}{2}\begin{bmatrix} (I)_A \otimes (I-\sigma_x)_B \otimes (I)_C - (\sigma_x)_A \otimes (I-\sigma_x)_B \otimes (\sigma_x)_C \\ -(\sigma_y)_A \otimes (I+\sigma_x)_B \otimes (\sigma_y)_C + (\sigma_z)_A \otimes (I-\sigma_x)_B \otimes (\sigma_z)_C \end{bmatrix} \\
&+\frac{1}{2}\begin{bmatrix} (I)_A \otimes (I+\sigma_y)_B \otimes (I)_C - (\sigma_x)_A \otimes (I+\sigma_y)_B \otimes (\sigma_y)_C \\ -(\sigma_y)_A \otimes (I+\sigma_y)_B \otimes (\sigma_x)_C + (\sigma_z)_A \otimes (I+\sigma_x)_B \otimes (\sigma_z)_C \end{bmatrix} \\
&\frac{1}{2}\begin{bmatrix} (I)_A \otimes (I-\sigma_y)_B \otimes (I)_C + (\sigma_x)_A \otimes (I-\sigma_y)_B \otimes (\sigma_y)_C \\ +(\sigma_y)_A \otimes (I-\sigma_y)_B \otimes (\sigma_x)_C + (\sigma_z)_A \otimes (I+\sigma_x)_B \otimes (\sigma_z)_C \end{bmatrix} \\
&-2(I)_A \otimes (I)_B \otimes (I)_C
\end{aligned}
\quad ; \quad \text{(B8)}
$$

$$
R_3 = 8\rho_{A|B|C} - 2(I)_A \otimes (I)_B \otimes (I)_C \quad . \qquad \text{(B9)}
$$

$$8\rho_4 = \frac{1}{2}\left[(I+\sigma_z)_A \otimes (I+\sigma_z)_B \otimes (I+\sigma_z)_C + (I-\sigma_z)_A \otimes (I-\sigma_z)_B \otimes (I-\sigma_z)_C\right] \quad . \qquad \text{(B10)}$$

$$R_4 = 8\rho_4 - (I)_A \otimes (I)_B \otimes (I)_C \ . \qquad \text{(B11)}$$

One should notice that we used un-normalized density matrices but one can consider them as contribute of normalized density matrices multiplied by some coefficients. Also while $\rho_1, \rho_2, \rho_3$ are considered as bi-separable density matrices $\rho_4$ is fully separable density matrix.

We find after lengthy but straightforward calculation that $R$ of Eqs. (B2, B3) is given by

$$R = 1/3(R_1 + R_2 + R_3 + R_4) = \frac{8(\rho_1 + \rho_2 + \rho_3 + \rho_4)}{3} - \frac{7}{3}(I)_A \otimes (I)_B \otimes (I)_C \quad . \qquad \text{(B12)}$$

Inserting Eq. (B12) into Eq. (B2) we get

$$8\rho_{WN} = (I)_A \otimes (I)_B \otimes (I)_C + p\left[\frac{8(\rho_1 + \rho_2 + \rho_3 + \rho_4)}{3} - \frac{7}{3}(I)_A \otimes (I)_B \otimes (I)_C\right] \quad . \qquad \text{(B13)}$$

We find that for $p \le 3/7$ the $|GHZ,3\rangle$ state mixed with white noise is explicitly bi-separable since only under this condition the coefficient of $(I)_A \otimes (I)_B \otimes (I)_C$ in Eq. (B13) will be non-negative. On the other hand we found by the use of the entanglement witness that for $p > 3/7$ this mixed state is genuinely entangled. The value $p = 3/7$ is the critical value for transition from genuine entanglement to bi-separability. The criterion $p > 3/7$ for genuine entanglement of the $|GHZ,3\rangle$ state mixed with white noise was derived previously [20] but here by explicit calculation of bi-separability we showed that this condition is indeed optimal.